**Validity Verification of the New TOEFL Writing Task Based on Classical Test Theory**

Zhang Yinyu

School of Foreign Language, Zhongnan University of Economics and Law

August 25, 2025

# Abstract

The TOEFL iBT has introduced the Academic Discussion Task (ADT) to assess test-takers' ability to engage in academic discourse, reflecting the growing emphasis on interactive communication skills in higher education. However, research on the ADT's validity and fairness particularly for culturally and linguistically diverse groups, such as Chinese students, remains limited. This study addresses this gap by employing Classical Test Theory (CTT) to evaluate the psychometric properties of the ADT among Chinese university students. This study finds a robust correlation between the ADT and the CET-6 writing and translation subscores. In addition, there is a high level of expert agreement regarding the construct validity evidence and the appropriateness of the scoring rubric. Furthermore, the results indicate that gender differences in validity indices are minimal. Taken together, these results suggest that the ADT is a valid measure for Chinese test-takers without gender discrimination. However, it is recommended that the cultural sensitivity of the scoring rubric be further refined and that the CET-6 subscores for writing be retained for predictive purposes, in order to better accommodate the needs of diverse test-taker populations. By addressing these issues, this study contributes to the broader discourse on fairness and validity in high-stakes language assessments.

**Validity Verification of the New TOEFL Writing Task Based on Classical Test Theory**

The TOEFL iBT (Internet-Based Test) is a comprehensive English language proficiency test that evaluates non-native speakers' ability to communicate effectively in academic settings. Since July 2023, the writing section has been revised to include two tasks: the Integrated Writing Task and the new Academic Discussion Task (ADT). The ADT simulates an online classroom discussion, requiring test-takers to present their views and provide arguments on a given topic within 10 minutes. However, few studies have examined its validity and fairness. Therefore, validation is essential to ensure that the ADT measures the intended construct and functions fairly across genders.

**Validity verification**

Classical Test Theory (CTT), a foundational theory in psychological and educational measurement, is widely used for analyzing the reliability and validity of standardized tests due to its operational simplicity and intuitive interpretation (Bachman, 1990; Swaminathan, n.d.). The core assumption of CTT is that each observed score (X) consists of a true score (T) and an error component (E), expressed as $X=T+E$ (Lord & Novick, 1968). Key propositions of CTT include reliability, validity, and item analysis. Reliability reflects the stability and consistency of measurement results, with commonly used indices such as Cronbach's alpha and split-half reliability. Validity concerns whether a test measures the intended psychological or ability constructs, with common types including content validity, construct validity, and criterion validity (Messick, 1989). In addition, CTT examines item difficulty and discrimination, which help evaluate the quality of test items and the overall measurement effectiveness (AERA/APA/NCME, 2014).

Within CTT, criterion validity is typically evaluated by correlating test scores with external criteria, such as standardized proficiency tests or academic outcomes. The Pearson correlation coefficient ($r$) quantifies the strength and direction of the linear relationship. Rather than invoking a universal cut-off, researchers interpret r with respect to decision stakes and domain norms. In high-stakes language testing, coefficients $\geq 0.60$ are typically deemed acceptable, whereas $\geq 0.80$ is preferred when serious consequences are attached to decisions (AERA/APA/NCME, 2014). Confidence intervals and sample sizes should accompany r to support accurate interpretation.

In CTT, content validity is primarily established through systematic expert judgment. Typically, a panel of subject-matter experts is invited to independently evaluate each test item for its relevance and representativeness in relation to the intended construct. A common approach is to use a Likert-type scale, typically ranging from 4 to 7 points, to rate each item (AERA/APA/NCME, 2014). Several quantitative indices have been developed to summarize expert ratings. The Item-Level Content Validity Index (I-CVI) is widely calculated as the proportion of experts who rate an item relevant (typically the top two points on a 4-point scale). Lynn (1986) recommends an I-CVI of $\geq 0.78$ for panels of six or more experts, and $\geq 0.83$ for five or fewer experts. At the scale level, the Scale-Level Content Validity Index (S-CVI) can be computed in two ways: the universal agreement approach (S-CVI/UA), which is the proportion of items that all experts rate as relevant, and the average approach (S-CVI/Ave), which is the average of the I-CVI values across all items. Recent guidelines suggest that an S-CVI/Ave of 0.80 or above indicates satisfactory content validity (Polit & Beck, 2006; AERA/APA/NCME, 2014).

In addition to these indices, to ensure scoring reliability,inter-rater reliability measures such as the Intraclass Correlation Coefficient (ICC) are increasingly used to assess the consistency of expert ratings. The ICC quantifies the degree of agreement among raters, with high ICC value indicating a high level of agreement between raters, demonstrating the scores' reliability (Shrout & Fleiss, 1979). Usually, values above 0.75 indicate good agreement and values above 0.90 indicate excellent agreement (Koo & Li, 2016). Descriptive statistics, including the mean and standard deviation of expert ratings, are also reported to reflect the central tendency and dispersion of expert judgments.

To further enhance content validity, iterative procedures such as Delphi rounds or focus group discussions may be employed. These methods allow for the refinement of item wording and ensure comprehensive coverage of the target construct through multiple rounds of expert feedback (AERA/APA/NCME, 2014).

Overall, the combination of expert panel ratings, quantitative indices (such as I-CVI, S-CVI, and ICC), and descriptive statistics provides a robust methodological foundation for evaluating content validity within the CTT framework. These methods and parameters offer clear criteria for interpreting the adequacy of test items and the consensus among experts, thereby supporting the validity argument for the assessment instrument. Together, these meathods provide a rigorous framework for validating language assessments under CTT, ensuring that tests are both empirically defensible and theoretically coherent.

**Validity Verification of TOEFL iBT**

Within the various sections of the TOEFL test, CTT has mainly been applied to the study of measurement properties in the reading, listening, and speaking sections.

For example, Sawaki et al. (2009) employed CTT reliability indices (Cronbach's α and subtest correlations) to examine the TOEFL iBT reading and listening sections, demonstrating that these traditional statistics effectively reveal measurement stability. Notably, their validation focused solely on the receptive skills, underscoring the broader pattern that CTT-based inquiries into TOEFL iBT have largely concentrated on reading and listening tasks rather than on writing performance. Brown and Ross (1996) applied CTT indices to evaluate the decision dependability of the TOEFL reading and listening sub-tests as well as the overall battery, yet no parallel CTT-driven study was conducted for the writing section. Lee (2006) employed generalizability theory, conceptually rooted in CTT, to investigate rater consistency and score reliability for TOEFL speaking tasks that integrated independent and integrated formats but again excluded writing tasks from the CTT framework. Along the same line, In'nami and Koizumi (2016) conducted a meta-analysis on 44 generalizability studies that primarily focused on speaking and writing tasks. They highlighted the relative scarcity of generalizability–and thus CTT-driven–investigations of writing, especially in large-scale assessment contexts. These studies demonstrate that CTT is widely used in the non-writing sections of the TOEFL, such as reading and listening, but its application to the writing section is relatively limited. This gap highlights the need for further research on the measurement properties of the TOEFL writing section.

**Validity Verification of TOEFL Writing Section**

The writing section of the TOEFL, due to its inherent subjectivity and complexity, has long been a focal point in language testing research. Traditionally, this section comprises the Independent Writing Task and the Integrated Writing Task, both of which have been extensively examined in terms of validity, scoring reliability, and fairness across diverse cultural and linguistic backgrounds (Cumming, 2013; Kunnan, 2004). Messick (1989) argued that the validity of writing tests encompasses not only content representativeness but also the fairness of scoring standards and the interpretability of results, advocating a multifaceted approach to validity evidence that includes both qualitative and quantitative analyses.

Empirical research has adopted a variety of methods to systematically investigate the validity of TOEFL writing. Chapelle et al. (2008) employed a comprehensive validity argument framework to validate the integrated writing task, emphasizing its ability to assess complex skills, such as synthesizing information from multiple sources, and highlighting the alignment between task design and the construct of academic English proficiency. Sawaki et al. (2009) used confirmatory factor analysis on large-scale test-taker data to examine whether the four TOEFL iBT skills (reading, listening, speaking, and writing) measured distinct language ability constructs. Their results showed that a four-factor model fit the data better than alternative models, supporting the test's validity in measuring separate language abilities. While these studies have provided a solid empirical foundation for TOEFL writing validity, few have applied CTT. According to Plakans and Gebril (2012), fewer than ten CTT-oriented studies had examined integrated second-language writing tasks as of 2012, highlighting the empirical gap in this domain. Moreover, most of these studies have focused on test-takers from North America and Europe, with relatively limited attention to Chinese test-takers, who represent the largest group of English learners worldwide. Alithough Cumming et al. (2005) analyzed discourse features in prototype TOEFL writing tasks based on Asian, they drew their samples primarily from Japanese and Korean candidates, neglecting Chinese examinees.

**Validity Verification of ADT**

The Academic Discussion Task (ADT) is designed to align more closely with the communicative demands of contemporary academic environments, where collaborative discussion and critical engagement are central (Educational Testing Service, 2023). Unlike the previous independent writing task, the ADT requires test-takers to respond to both a prompt and simulated peer contributions, thereby emphasizing skills such as synthesizing multiple perspectives, constructing coherent arguments, and engaging in interactive written discourse (Wang, 2024). This shift reflects a broader trend in language assessment toward tasks that more closely mirror authentic academic practices. However, as noted by Davis and Norris (2023), the increased interactivity and contextualization of the ADT may introduce new sources of construct-irrelevant variance, such as differential familiarity with online discussion conventions.

Despite its theoretical advantages, current scholarship indicates that empirical scrutiny of the ADT remains scarce, particularly with respect to its measurement properties, such as validity and fairness. For example, Cushing (2025) explicitly notes that the 2023 revision replaced the traditional independent essay with the 10-minute Academic Discussion task, yet the report limits itself to task description and scoring logistics; no systematic validity evidence is provided. Similar to the research on the validity of TOEFL writing, systematic studies on the ADT—particularly among Chinese test-takers–also remain scarce.

**Research Gap**

A review of the current literature highlights several important gaps in the empirical study of the newly introduced Academic Discussion Task (ADT). While previous research has provided valuable insights into the validity, reliability, and fairness of the traditional Independent and Integrated Writing Tasks, systematic investigations of the ADT remain limited. Most existing studies have focused on describing the design and scoring procedures of the ADT, with little attention given to its measurement properties, such as construct validity and fairness.

Moreover, the majority of research on TOEFL writing has concentrated on test-taker populations from North America, Europe, and a few Asian countries, such as Japan and Korea. Studies specifically examining Chinese test-takers, who constitute one of the largest groups of TOEFL examinees, are relatively scarce. Given the unique linguistic and cultural characteristics of Chinese learners, it is crucial to determine whether the ADT accurately assesses their academic discussion skills and whether it operates fairly across gender. In summary, there is a clear need for empirical research that evaluates the validity and fairness of the ADT among Chinese test-takers.

**The current study**

Therefore, to address these research gaps, the present study adopts Classical Test Theory (CTT) as its framework to systematically analyze the psychometric properties of the ADT among Chinese university students. Specifically, it focuses on the validity and fairness of the ADT across different subgroups. The research questions are as follows:

1) Does the ADT effectively evaluate the intended construct?

2) Does the ADT discriminate against test-takers of different genders?

## Methodology

The current study examines the validity of the ADT via criterion validity and content validity. Detailed information is presented as follows.

### Participants

For criterion validity, 300 Chinese university students participated in the current study, 43% of whom were male. Most participants were in their third or fourth year of undergraduate study. All participants had passed the College English Test Band 6 (CET-6), with an average overall score of 502 and an average writing and translation score of 138. For content validity, five full professors from Chinese universities, all holding doctoral degrees and at least five years of experience in English academic writing instruction or assessment, volunteered to serve as expert raters. Informed consent was obtained via email.

### Instruments

#### *Test Item*

The ADT item used in this study was taken from the TOEFL test administered in December 2023. It simulated a sociology class discussion on population trends in urban and rural areas. The test item is as follows.

*Dr. Achebe asked how governments could encourage people to live in rural areas. Claire suggested increasing financial support for farmers to offset the high costs of starting an agricultural career. In contrast, Kelly emphasized the greater amenities in urban centers*

*and stated she would consider rural living only if more entertainment and cultural businesses were available there. The task required students to analyze the factors influencing people's preferences for urban or rural living and to propose effective governmental strategies to attract residents to rural areas, with an effective response expected to be at least 100 words.*

The evaluation framework was based on the official TOEFL scoring standards (ETS, 2023). The scoring criteria assessed content, development, organization, language use, vocabulary, grammar, and mechanics. Each response was scored on a scale of 0 to 5 by two independent raters. Scores were then averaged to produce a raw score between 0 and 10, which was subsequently converted to a scaled score between 0 and 30.

### *Questionnaires*

To evaluate both the construct coverage and the scoring appropriateness of the ADT, two concise seven-point Likert questionnaires including the Construct Questionnaire and the Scoring Rubric Questionnaire. They were designed and administered via Wenjuanxing, a widely used Chinese platform for collecting survey data. The Construct Questionnaire asked each expert to rate, on a scale from 1 (not at all) to 7 (fully), the extent to which the ADT addresses four constructs listed on the TOEFL website: critical thinking, academic writing skills (eg., vocabulary, grammar, and conventions), argumentation (eg., logical structure), and knowledge. Then the Scoring Rubric Questionnaire invited the same experts to rate the ADT's official scoring rubric's capacity to assess and differentiate test-taker performance, again using the identical 1-to-7 scale.

**Procedures**

### *Criterion Validity*

Before administering the ADT task, participants' background information–including their English proficiency scores (e.g., CET-6) and academic backgrounds–was collected to ensure a representative sample. Participants were asked to complete the ADT task within 10 minutes, adhering to the official TOEFL iBT time limits for the writing section (Educational Testing Service, 2023).

### Content Validity

To ensure the validity and reliability of the two questionnaires, both instruments were released simultaneously, and all responses were collected anonymously within 48 hours and imported directly into RStudio (RStudio Team, 2025) for I-CVI, S-CVI, ICC and descriptive statistical analysis.

## Results

### Criterion Validity

This section utilizes Bayesian correlation analysis via JASP 0.19.3 (JASP Team, 2025) to investigate the relationship between the writing and translation subscore of the CET-6 (CET-6W+T), the total CET-6 score (CET-6S), and the TOEFL Academic Discussion Task (ADT). The analysis is structured into two main parts. First, it examines the correlations between these variables without considering gender, focusing on two distinct groups: the subscore group (CET-6W+T and TOEFL ADT) and the total score group (CET-6S and TOEFL ADT). In JASP, TOEFL ADT scores were matched separately with CET-6W+T and CET-6S. Each group is analyzed through Bayesian Pearson correlations, correlation pairwise scatter plots, and descriptive statistics. Subsequently, to explore potential measurement bias and ensure equitable validity, this study conducts a gender-based comparative analysis for both the subscore and total score groups. Between-group and gender-based comparisons elucidate differences in correlations between CET-6W+T, CET-6S and TOEFL ADT, as well as practical implications for language assessment.

### Subscore Group Analysis: CET-6W+T and TOEFL ADT

According to the descriptive statistics of CET-6W+T and TOEFL ADT, CET-6W+T exhibits a broader score distribution (range: 100–193) compared to TOEFL ADT (range: 16–28), with a significantly higher standard deviation (26.398 vs. 3.111) and mean (143.860 vs. 23.387).

The Bayesian Pearson Correlation between CET-6W+T and TOEFL ADT is extremely positive (Pearson's $r$=0.926, $p$<0.001). The effect size ($R^2$=0.857) indicates that the specialized subscore explains 85.7% of the variance in TOEFL writing performance. The Bayes Factor ($BF_{10}$) is $9.638 \times 10^{123}$, providing overwhelming evidence for the correlation. The 95% confidence interval for the correlation is [0.907, 0.940]. To visualize the relationship, Figure 1 presents the Bayesian Correlation Pairwise Scatter Plot of TOEFL ADT and CET-6W+T.

**Figure 1**

*Bayesian Correlation Pairwise Scater Plot of CET-6W+T and TOEFL ADT*

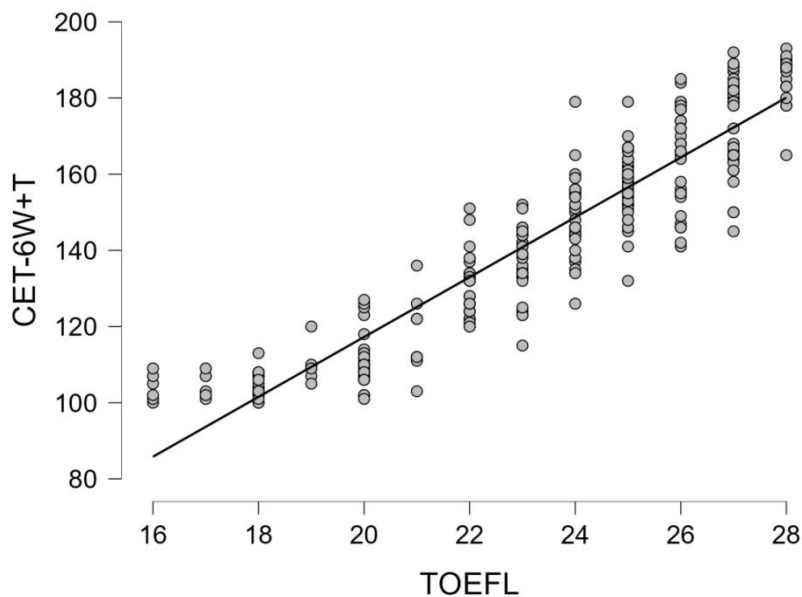

The scatter plot of TOEFL ADT versus CET-6W+T (TOEFL ADT: 16–28 vs. CET-6W+T: 100–193) shows tightly clustered data points along the regression line, indicating high consistency in skill evaluation. Its slope of approximately 8.2 demonstrates that a 1-point increase in TOEFL ADT corresponds to an 8.2-point rise in CET-6W+T, reflecting proportional scaling between the tests. Outliers are observed, such as a TOEFL ADT score of 28 paired with a CET-6W+T score of 150.

***Total Score Group Analysis: CET-6S and TOEFL ADT***

The following section probes into the relationship between CET-6S (CET-6 total score) and TOEFL ADT. JASP shows that the mean score for CET-6S is 516.64, with a standard deviation of 55.08, a minimum score of 393, and a maximum score of 648. While TOEFL ADT maintains the same mean of 23.387 and standard deviation of 3.111, ranging from 16 to 28.

To further investigate the association between CET-6S and TOEFL ADT, a Bayesian Pearson correlation analysis is conducted. The results indicate a moderate positive correlation ($r = 0.547$, $p < 0.001$), with CET-6S explaining 30.0% of the variance in TOEFL ADT writing ($R^2 = 0.300$). The Bayes Factor ($BF_{10}$) is $5.133 \times 10^{21}$. and the 95% confidence interval for the correlation is [0.460, 0.620]. Figure 2 displays the correlation between CET-6S and TOEFL ADT more directly and clearly.

**Figure 2**

*Bayesian Correlation Pairwise Scatter Plot of CET-6S and TOEFL ADT*

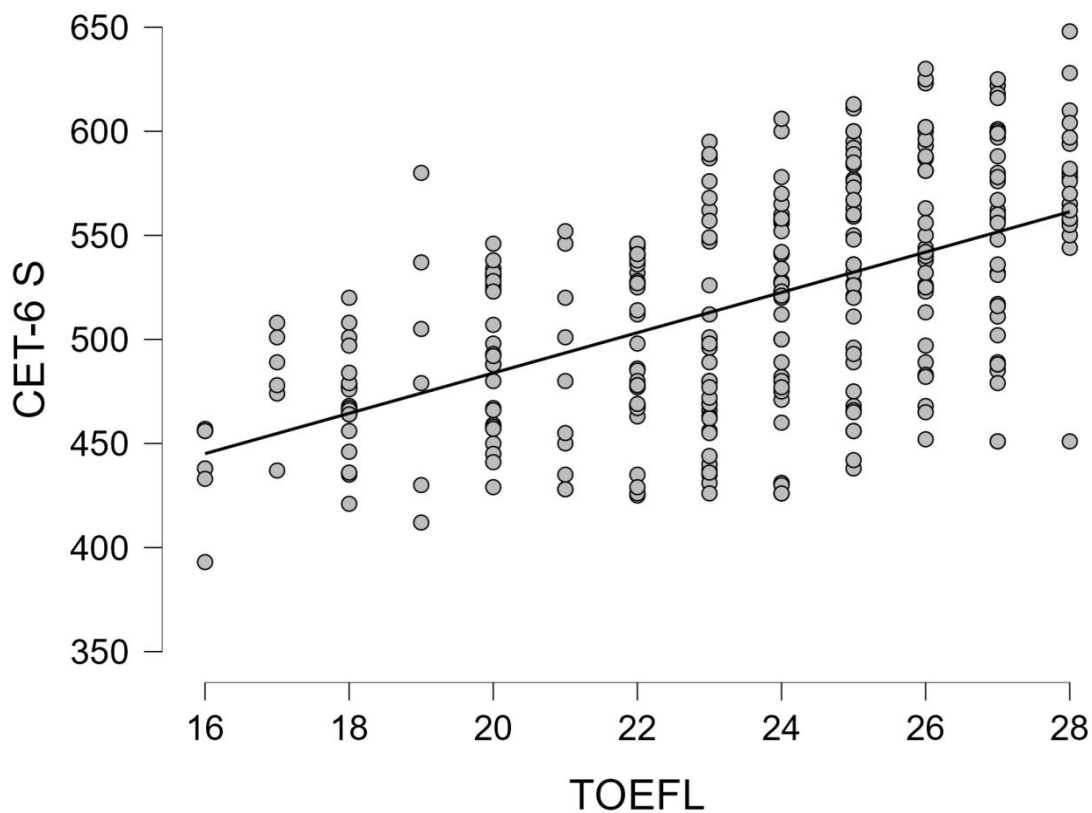

The scatter plot in Figure 2 (TOEFL ADT: 16–28 vs. CET-6S: 393–648) confirms these findings, with data points forming a dispersed "cloud-like" distribution, a shallower slope ($r = 0.547$), and higher variability (standard deviation = 55.083). Outliers, such as CET-6S = 600 paired with TOEFL ADT = 18, are presented.

***Between-Group Comparison***

The subscore group (CET-6W+T and TOEFL ADT) shows a much higher correlation ($r = 0.926$, $BF_{10} = 9.638 \times 10^{123}$) than the total score group (CET-6S and TOEFL ADT, $r = 0.547$, $BF_{10} = 5.133 \times 10^{21}$). The effect size for the subscore group is 185.7% higher than that for the total score group. The subscore group's data points are more densely clustered, and its standard deviation is lower (26.398 vs. 55.083).

All analyses described above are performed on the whole sample, and this invites a pivotal follow-up inquiry: Does TOEFL ADT's writing assessment demonstrate equitable validity across genders? To investigate potential measurement bias, the following section stratifies the Bayesian framework by gender. Using identical correlation models (JASP 0.19.3), it examines whether male (n=130) and female (n=170) subgroups exhibit systematically divergence in the subscore group (CET-6W+T and TOEFL ADT) and the total score group (CET-6S and TOEFL ADT). Significant differences between male and female in correlation strength (*r*) or explained variance ($R^2$) would indicate threats to test fairness, whereas consistent patterns would support the fairness of the ADT. Meanwhile, this section also explores other gender-specific differences in performance metrics.

### *Cross-gender Analysis: Relationship between CET-6W+T and TOEFL ADT (Subscore Group)*

This section presents a gender-based analysis of the relationship between the CET-6 writing and translation subscore (CET-6W+T) and TOEFL ADT performance. The analysis is based on a sample of 130 male and 170 female participants.

The results show that female participants achieved higher mean scores than males in both CET-6W+T and TOEFL ADT. Specifically, the mean CET-6W+T score is 146.92 for females and 139.85 for males, while the mean TOEFL ADT score is 23.78 for females and 22.88 for males, with decreases of approximately 3.3% and 3.0%, respectively. In terms of score distribution, the CET-6W+T scores range from 100 to 189 for males and from 100 to 193 for females, indicating a higher maximum score among females, while both genders shared the same score range (16–28) in TOEFL ADT.

The standard deviation of CET-6W+T scores and TOEFL ADT scores are slightly higher for females (CET-6W+T: 26.380; TOEFL ADT: 3.162) compared to males (CET-6W+T: 25.979; TOEFL ADT: 2.978), indicating a slightly greater variability in writing and translation performance among females. Turning to the correlation between the two assessments, both groups demonstrate extremely strong positive correlations between CET-6W+T and TOEFL ADT ($r > 0.91$), with robust evidence and a narrow 95% CI for males ($BF_{10} > 10^{56}$, 95% CI [0.910, 0.955]) and females ($BF_{10} > 10^{64}$, 95% CI [0.885, 0.936]). Specifically, the male group exhibits a slightly higher Pearson's $r$ (0.938 vs. 0.915) compared to the female group. Correspondingly, the effect size, as measured by $R^2$, indicates that CET-6W+T scores explain 87.98% of the variance in TOEFL ADT performance for males and 83.72% for females, representing a difference of 4.26 percentage points. Additionally, the Bayes Factors for both groups are substantially above 100, providing overwhelming evidence for the observed correlations.

To further visualize the relationship and provide concrete evidence for the comparison between the male and female groups, Figure 3 and Figure 4 are presented below.

**Figure 3**

*Bayesian Correlation Pairwise Scatter Plot of CET-6W+T and TOEFL ADT for Male Group*

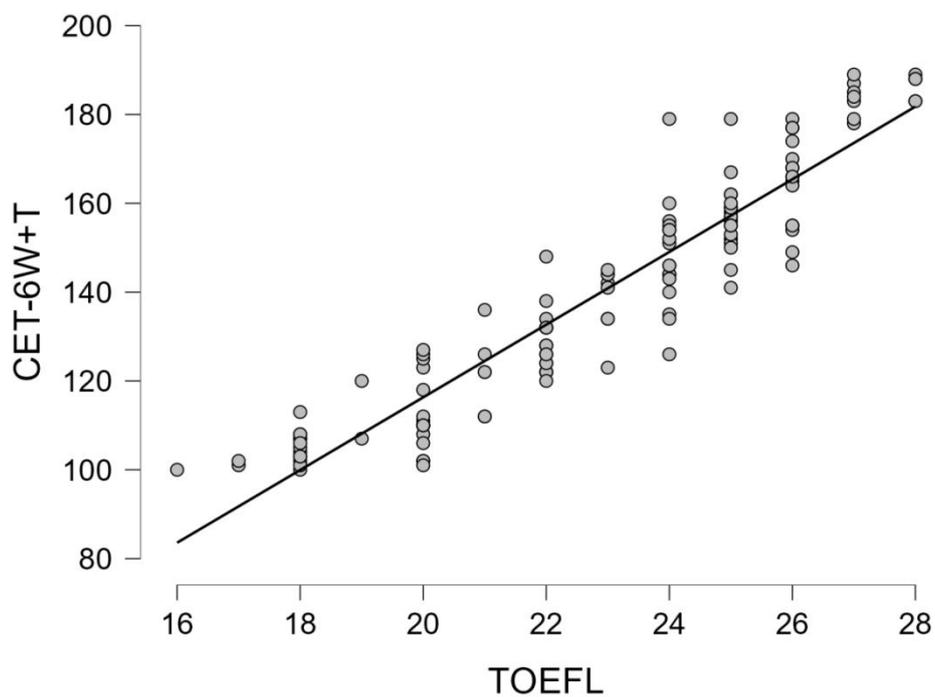

**Figure 4**

*Bayesian Correlation Pairwise Scatter Plot of CET-6W+T and TOEFL ADT for Female Group*

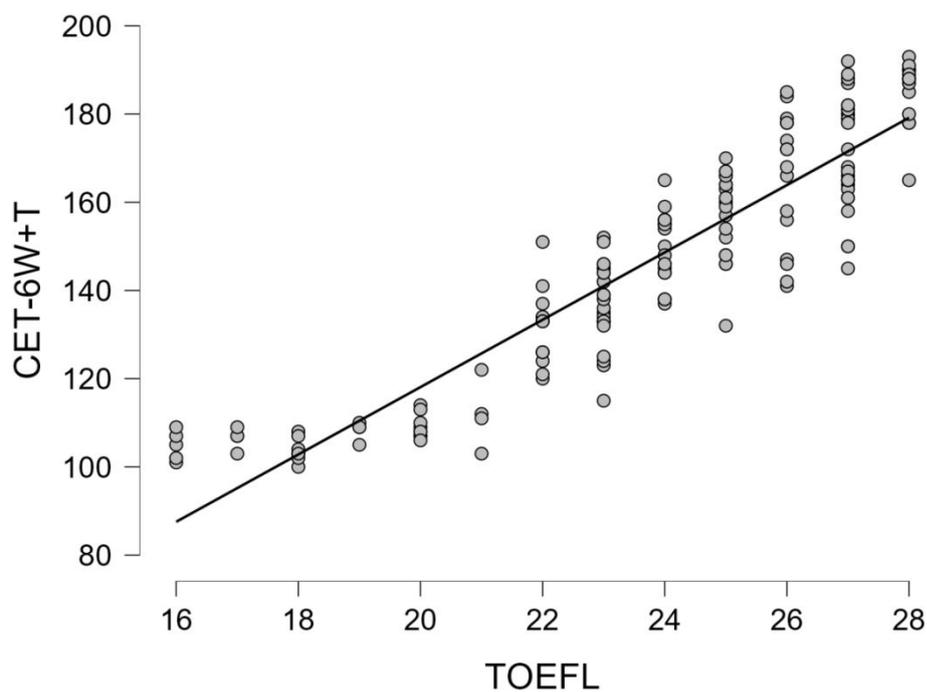

As can be seen from Figures 3 and 4, both the male and female groups' scatter plots reveal tightly clustered data points along the regression line, indicating a high correlation between CET-6W+T and TOEFL ADT across genders. For the male group, a 1-point increase in TOEFL ADT corresponds to an approximately 8.5-point rise in CET-6W+T, while for the female group, the corresponding increase is around 8.0 points. The slight differences in the slope and clustering density of the data points between the two groups are consistent with the minor differences in correlation coefficients and effect sizes mentioned earlier, further supporting the fairness of ADT.

***Gender-Based Analysis: Relationship between CET-6S and TOEFL ADT (Total Score Group)***

This section provides a gender-based analysis of the correlation between the CET-6 total score (CET-6S) and TOEFL ADT score.

The results indicate that the female group achieves both a higher mean and a higher maximum score in CET-6S than the male group, with an increase of approximately 1.1% in mean and an upper limit of 648 compared to 630 for males. The standard deviation of CET-6S scores is also higher for females (57.727) than for males (51.480), indicating greater variability in overall English proficiency among females.

In terms of correlation, the relationship between CET-6S and TOEFL ADT is moderate for both groups (r ≈ 0.55), substantially lower than the subscore group. Both genders show non-overlapping confidence intervals excluding zero and stable Bayes Factors ($BF_{10} > 10^8$ for males; $BF_{10} > 10^{11}$ for females), indicating consistent validity across gender. Specifically, the female group shows a slightly higher Pearson's r (0.551 vs. 0.539) and higher Bayes Factors (female: $1.091×10^{12}$; male: $2.760×10^{8}$), indicating that the female group demonstrates stronger correlation between CET-6S and TOEFL ADT than the male group. To further visualize the relationship and provide concrete evidence for the correlation comparison between the male and female groups, Figure 5 and Figure 6 are presented below.

**Figure 5**

*Bayesian Correlation Pairwise Scatter Plot of CET-6S and TOEFL ADT for Female Group*

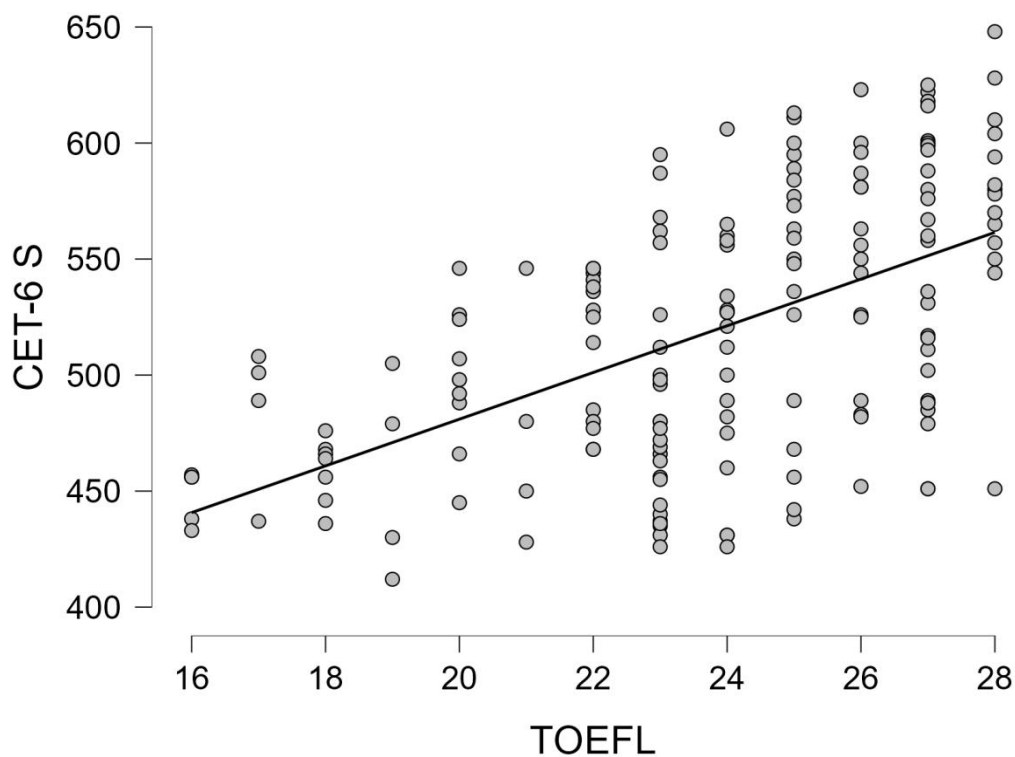

**Figure 6**

*Bayesian Correlation Pairwise Scatter Plot of CET-6S and TOEFL ADT for Male Group*

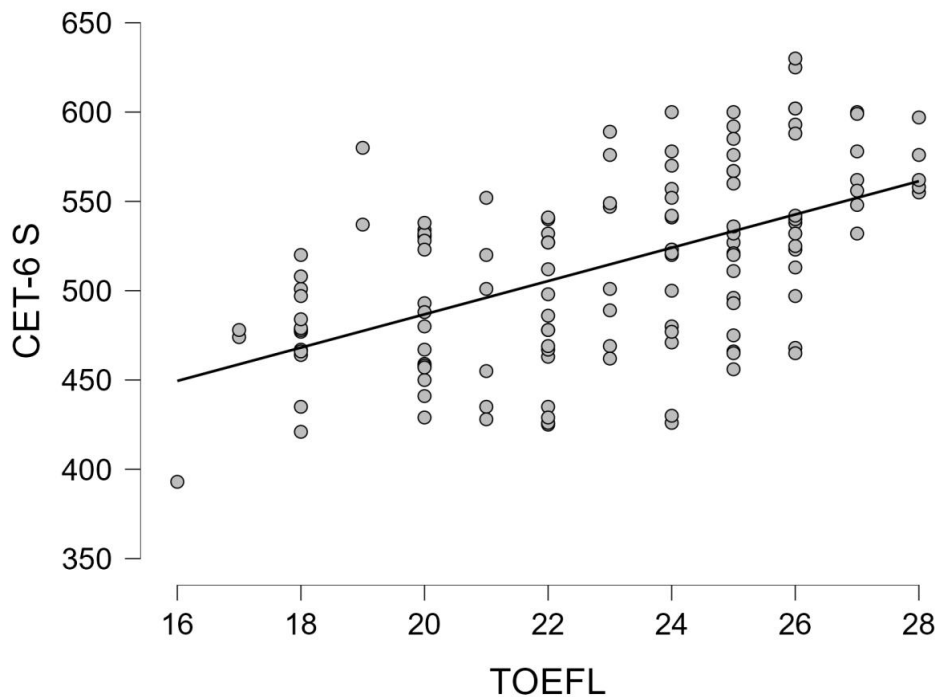

As shown in Figures 5 and 6, the scatter plots for both male and female groups reveal a dispersed, "cloud-like" distribution of data points. This pattern indicates a more variable relationship between CET-6S and TOEFL ADT compared to the subscore group.

In conclusion, the data analysis, encompassing both overall and gender-based comparative analyses, offers a detailed examination of the predictive validity of CET-6 scores for TOEFL ADT performance. The overall findings demonstrate that CET-6W+T is a highly reliable predictor of TOEFL ADT performance, exhibiting both strong correlation and large effect size. In contrast, the total CET-6 score (CET-6S) shows a weaker correlation due to its composite nature. The gender-based comparative analysis shows subtle differences across genders in the observed relationships.

**Content Validity**

To verify the content validity of the new TOEFL writing task, Academic Discussion Task (ADT), a multi-faceted expert review is conducted, assessing item-construct alignment and scoring rubric appropriateness.

### Task-construct Alignment

Five subject-matter experts rate the extent to which ADT measures four target constructs (e.g., argument integration, academic discourse) on a 7-point Likert scale (1 = not covered, 7 = fully covered). To evaluate the consistency of their ratings, an Intraclass Correlation Coefficient (ICC) analysis was performed using RStudio. The results indicate that the single-rater ICC ($ICC_2$) is 0.44 (95% CI [0.08, 0.93], p = 0.002), whereas the average-rater ICC ($ICC_2k$) reached 0.80 (95% CI [0.31, 0.98]). Additionally, the I-CVI for construct-coverage items = [0.80, 1.00, 1.00, 1.00], demonstrating that all items except Q1 all exceed the "good" threshold of 0.83, but the I-CVI of Q1 is still relatively high and acceptable (Lynn, 1986). While the S-CVI/Ave = 0.95 and the S-CVI/UA = 0.80, meeting Polit & Beck's (2006) excellence level.

### Scoring Efficacy

Regarding scoring efficacy, the same professors evaluated the appropriateness of the ADT scoring criteria on a 7-point Likert scale (1 = not effective, 7 = very effective). Descriptive statistics, obtained via RStudio, show that the average rating was 5.2 , significantly exceeding the scale midpoint of 4 (t = 12.1, p < 0.001) with minimal dispersion (SD = 0.45), along with the I-CVI for rubric appropriateness = 1.00.

### Overall Evaluation

The robust item-construct alignment ($ICC_2k = 0.80$, S-CVI/Ave = 0.95, S-CVI/UA = 0.80) and low SD ensure the reliability and high rubric appropriateness (M = 5.2) guaranteeing scoring fidelity, provide evidence that the new TOEFL writing task, ADT, has acceptable content validity in the current sample.

## Discussion

### Validity Verification of ADT

One of the core findings regarding the ADT's validity is its exceptionally high correlation with the CET-6 writing and translation subscore. This provides strong statistical support for the relationship (Kass & Raftery, 1995).. This alignment likely originates from overlapping scoring criteria between the two tests, such as an emphasis on explanation, coherence, grammatical accuracy, and linguistic complexity (Educational Testing Service, 2024; National Education Examinations Authority, 2016), enhancing the consistency of assessing specialized writing skills. The stability of the correlation further reinforces its criterion validity. For the outliers in Figure 1, they may reflect divergent test strategies. For instance, Zhang (2022) contrasts TOEFL's emphasis on integrated academic skills with CET-6's focus on basic vocabulary and grammatical accuracy, illustrating how test-specific focus can lead to skill mismatches and differential candidate performance. As a result, candidates excelling in one test's focus may underperform in the other due to skill mismatches, warranting qualitative investigation. Moreover, compared with ADT, CET-6W+T shows a wider spread of scores and a higher average level. This wider variability likely stems from two factors: 1) the combined structure of writing and translation, where strengths in one task can offset weaknesses in the other, and 2) the broader scoring scale, which allows scores to spread more widely.

Although the effect size is moderate and weaker than that of the subscore group, the finding remains robust across analytic approaches. These results support ADT's criterion validity while indicating that total scores introduce more noise when predicting writing ability. This attenuation likely arises from CET-6S incorporating more non-writing modules such as listening and reading, which reduce its specificity in predicting writing proficiency as students might excel in listening or reading but lack essential writing strategies, or they might achieve high writing scores while underperforming in other modules. This result aligns with and extends prior findings on TOEFL writing research. Sawaki et al. (2009) showed that TOEFL iBT skills (listening, speaking, reading, and writing) represent distinct factors, confirming writing as a separable construct. This is consistent with our finding that CET-6W+T—a writing-focused measure—correlates more strongly with ADT than CET-6S, which is a broader proficiency measure. Moreover, it also aligns with Messick's (1989) "construct contamination theory," which argues that composite scores influenced by multiple abilities diminish the validity of predictions for a single skill. This underscores ADT's strength in targeting writing-specific constructs while highlighting the reduced construct purity of multi-skill measures.Different from the outliers in Figure 1, those in Figure 2 demonstrate the divergent test focus: comprehensive and specialized competencies. In addition, Figure 2 and the standard deviation for CET-6S indicates a relatively wide spread of total scores around the mean. This spread may also be attributed to the comprehensive nature of the CET-6S, which encompasses multiple sections such as listening, reading, writing, and translation. The aggregation of diverse skills into a single score dilutes domain-specific precision, as high performance in one module (e.g., listening) may offset weaknesses in others (e.g., writing). When closely comparing the subscore group with the total score group, the validity advantages of subscores become more evident. The subscore group's effect size is substantially larger, and its Bayes Factor far exceeds that of the total score group, providing overwhelming support for the strong correlation between CET-6W+T and TOEFL ADT (Wagenmakers et al., 2018) and the superior precision in predicting writing ability (Bachman & Palmer, 2010). The scatter plots also illustrate this contrast: subscores align more closely with ADT, while total scores appear more dispersed, highlighting their limitations. This comparison underscores the importance of considering subscores for a more nuanced evaluation of specific proficiency and highlights the need for balanced ability distribution in test design. Consequently, ADT's strong correlation with CET-6W+T and the moderate correlation with CET-6S support its criterion validity

As for content validity, expert evaluations provide strong support for ADT. The high I-CVI for each item and the strong S-CVI for the Construct Questionnaire indicate close task–construct alignment, and the high $ICC_2k$ further supports the reliability of these results. Similarly, the Scoring Rubric Questionnaire showed high I-CVI values and strong agreement among experts regarding the appropriateness of the scoring criteria. The low SD across expert ratings further indicates consistency and high reliability.

In summary, ADT demonstrates robust criterion and content validity.

**Evaluating the Gender Fairness of ADT**

In terms of fairness, cross-gender analyses in this study revealed that the higher mean scores in CET-6W+T, CET-6S, and TOEFL ADT, along with the higher score ceiling in CET-6W+T and CET-6S, suggest that females tend to perform more strongly in writing and translation, which is consistent with documented female advantages in language expression and attention to detail (Halpern, 2012). Females' higher mean and score ceiling in CET-6S indicate stronger overall English proficiency, especially at the top end of performance, which is consistent with previous research on female superiority in language tasks (Hyde & Linn, 1988; Reilly et al., 2019). The smaller gender gap regarding the mean observed in TOEFL ADT, along with the identical and relatively narrow score range for both groups, may be attributed to the fact that the task exclusively assesses writing, thereby reflecting writing competence without interference from other sections (e.g., listening, reading, or translation). Moreover, the narrower score range of TOEFL ADT, compared with CET-6W+T and CET-6S, may help explain the smaller observed gender differences, as ADT provides fewer opportunities to detect variation at the extremes. In contrast, broader score range allows for greater observation of gender differences at the extremes such as CET-6S. The wider score variation among females may reflect more diverse writing strategies, leading to a broader range of outcomes. However, these differences are still subtle,for both genders suggesting that CET-6W+T and TOEFL ADT effectively ensure the fairness across genders.

In terms of correlations, the slightly higher coefficients and explanatory power observed in the male group's subgroup suggest a more consistent relationship between their CET-6W+T and TOEFL ADT performance. This consistency may be attributed to males' potential inclination towards structured, rule-based writing strategies in standardized tests, which can enhance consistency in writing performance across different examinations (Graham & Perin, 2007). Meanwhile, although females generally outperform males in language-related tasks (Hyde & Linn, 1988; Reilly et al., 2019), their potentially more diverse writing strategies across different exams (Ellis et al., 2023) may contribute to greater score variability and, consequently, slightly lower correlations. However, it is important to note that the gender differences in correlation coefficients and effect sizes in ADT remain relatively small, indicating that TOEFL ADT is largely fair to candidates of different genders. Furthermore, according to Figure 3 and 4, the consistent pattern across both groups suggests that the relationship between CET-6W+T and TOEFL ADT holds reliably across genders

In contrast, females show a slightly higher correlation between CET-6S and ADT. The slightly higher correlation and explanatory power in the female group might attributed to famale's more balanced distribution of abilities across modules or a stronger link between overall English proficiency and writing skills among females (Hyde, 2005). Nevertheless, the overlapping score ranges and small gender differences in correlations suggest that the relationship between CET-6S and TOEFL ADT is largely consistent across groups, despite the slight advantage observed among females. Additionally, expert approval of the scoring rubric is high, further supporting the fairness of the ADT scoring process.

ADT's fairness aligns with findings on traditional TOEFL tasks. Kunnan (2004) observed small to moderate gender differences in TOEFL CBT writing (Cohen's d≈0.25 for independent tasks, d≈0.10 for integrated tasks), with females scoring higher which is consistent with our finding of slight female's advantages in ADT. Reilly et al. (2019) noted larger gender differences in writing than in reading or listening, possibly due to writing's subjectivity, but ADT reduces this risk by using standardized input, much like integrated writing tasks that limit the influence of background knowledge (Plakans & Gebril, 2012).

Together, these findings reinforce ADT's fairness, indicating that its design effectively balances potential gender-related variability in writing performance.

**ADT's Strength**

Compared with traditional TOEFL writing tasks, ADT possesses unique strengths that enhance its assessment value. Plakans and Gebril (2012) found that TOEFL integrated writing emphasizes multi-source integration, while independent writing focuses on personal viewpoint development which is echoed by Cumming et al. (2005). ADT advances this approach by introducing authentic academic discussion scenarios that require test-takers to integrate standardized peer perspectives. This design retains integrated writing's information integration strength while enhancing authenticity and interactivity (Wang, 2024), allowing for better reflection of the communicative demands of real-world academic contexts. Moreover, according to Swain's (1985) Output Hypothesis, which emphasizes the importance of meaningful language production for second language development, the necessity to respond to and integrate multiple perspectives encourages deeper cognitive processing and helps EFL learners internalize complex language structures. Its interactive format also supports the development of higher order thinking and discourse comprehension. Drawing on Vygotsky's (1978) sociocultural theory, the task's collaborative nature fosters social interaction and negotiation of meaning, both of which are essential for cognitive and linguistic growth. By exposing test-takers to diverse argumentation strategies and linguistic forms, ADT promotes the ability to synthesize information and construct well-reasoned academic arguments.

Second, in terms of assessment dimensions, ADT goes beyond evaluating basic skills such as grammar, which are commonly assessed in traditional writing tasks. It further incorporates the evaluation of academic interaction, thereby broadening the scope of writing assessment (Chapelle et al., 2008). Third, ADT features increased specificity in measuring writing ability. By using standardized input including peer viewpoints and discussion contexts effectively eliminates performance differences stemming from varying background knowledge or personal experiences. This design mirrors the standardized materials used in integrated writing tasks but with a more focused emphasis on writing ability, ensuring that all test-takers face equivalent information conditions. Additionally, it exclude non-writing skills such as listening from the assessment, ADT reduces construct-irrelevant variance, isolating writing ability as the sole focus of evaluation. This targeted approach minimizes the influence of unrelated competencies, preventing them from disproportionately affecting scores and thus strengthening fairness and improving the interpretability of results (Plakans & Gebril, 2012). As a result, ADT is more advanced than the traditional TOEFL writing tasks.

## Conclusion

This study evaluated the validity and fairness of the TOEFL Academic Discussion Task (ADT) among Chinese university students, the findings are as follows. First, regarding validity, ADT demonstrated strong criterion validity, with a very high correlation to CET-6 writing and translation subscore, confirming its effectiveness in assessing specialized writing skills. In contrast, its moderate correlation with the CET-6 total score suggests that ADT is more narrowly focused on writing proficiency rather than overall language ability. Expert evaluations further validated its content validity. Second, in terms of fairness, gender-based analyses revealed minimal gender differences in ADT. Though females scored slightly higher on average, score ranges overlapped substantially, and variability in performance was comparable across genders. These results indicate that ADT is free from systematic gender bias.

This study offers important implications for second language acquisition, language teaching, and English testing. For second language acquisition and teaching, the high validity of ADT confirms that academic discussion skills–including integrating peer perspectives, constructing coherent arguments, and engaging in interactive discourse–are measurable and critical components of academic English proficiency. This highlights the need for teaching practices to emphasize these skills, moving beyond traditional writing drills to include authentic discussion-based activities. For English testing, the findings provide empirical evidence supporting the psychometric quality of ADT, particularly its ability to assess writing-specific constructs more precisely than comprehensive test scores. This validates the TOEFL's revision to include ADT and suggests that using writing subscores (e.g., CET-6W+T) as predictive indicators can enhance the accuracy of specific proficiency evaluations. Additionally, by focusing on Chinese test-takers–one of the largest TOEFL populations–this study fills a gap in existing research and provides cross-cultural insights into the validity and fairness of ADT. It underscores the importance of context-specific validation in high-stakes language assessments and provides a foundation for future studies on diverse test-taker groups.

However, this study also has several limitations. First, the sample was restricted to Chinese university students who had passed CET-6, limiting generalizability to other populations, such as lower-proficiency learners or non-university test-takers. Second, the research methodology relied solely on CTT, which focuses on observed scores and may not capture more nuanced psychometric properties. Third, the investigation of fairness was limited to gender, neglecting other potential sources of variance, such as regional differences in English education or prior experience with ADT, which may affect ADT performance. Finally, this study employed only one ADT task item (a sociology-related topic) in the questionnaires, so the results may not fully capture the task's validity across diverse academic domains.

In conclusion, ADT is a valid and fair tool for assessing academic writing skills among Chinese students, with implications for both language teaching and test development. Addressing its limitations in future research will further strengthen its utility in global language assessment.